\documentclass[twocolumn,superscriptaddress,aps,preprintnumbers,amsmath,amssymb,prl,nofootinbib]{revtex4}

\usepackage{epsfig}
\usepackage{amsmath}
\usepackage{amssymb}
\usepackage{subfigure}
\usepackage{cancel}
\usepackage{textcomp}
\usepackage{calc}
\usepackage{graphicx}
\usepackage{dcolumn}
\usepackage{color}
\usepackage{xcolor}
\usepackage{pifont}

\usepackage{hyperref}
\hypersetup{colorlinks=true,urlcolor=blue,linkcolor=red,citecolor=green!60!black}

\begin{document}

%%%%%%%%%%%%%%%%%%%%%%%%%%%%%%%%%%%%%%%%%%%%%%%%%%%%%%%%%%%%%%%%%%%%%%%%%%%%%%%%%%%%%%%%%%%%%%%%%%%%

\title{Probing the scale of grand unification with gravitational waves}

%%%%%%%%%%%%%%%%%%%%%%%%%%%%%%%%%%%%%%%%%%%%%%%%%%%%%%%%%%%%%%%%%%%%%%%%%%%%%%%%%%%%%%%%%%%%%%%%%%%%

\preprint{CERN-TH-2019-215}
\preprint{DESY 19-210}
\preprint{IPMU 19-0179}

%%%%%%%%%%%%%%%%%%%%%%%%%%%%%%%%%%%%%%%%%%%%%%%%%%%%%%%%%%%%%%%%%%%%%%%%%%%%%%%%%%%%%%%%%%%%%%%%%%%%

\author{Wilfried Buchmuller}
\email{wilfried.buchmueller@desy.de}
\affiliation{Deutsches Elektronen Synchrotron DESY, 22607 Hamburg, Germany}

\author{Valerie Domcke}
\email{valerie.domcke@desy.de}
\affiliation{Deutsches Elektronen Synchrotron DESY, 22607 Hamburg, Germany}

\author{Hitoshi Murayama}
\email{hitoshi@berkeley.edu}
\affiliation{Deutsches Elektronen Synchrotron DESY, 22607 Hamburg, Germany}
\affiliation{Department of Physics, University of California, Berkeley, CA 94720, USA}
\affiliation{Lawrence Berkeley National Laboratory, Berkeley, CA 94720, USA}
\affiliation{Kavli IPMU (WPI), University of Tokyo, Kashiwa 277-8583, Japan}

\author{Kai Schmitz}
\email{kai.schmitz@cern.ch}
\affiliation{Theoretical Physics Department, CERN, 1211 Geneva 23, Switzerland}

%%%%%%%%%%%%%%%%%%%%%%%%%%%%%%%%%%%%%%%%%%%%%%%%%%%%%%%%%%%%%%%%%%%%%%%%%%%%%%%%%%%%%%%%%%%%%%%%%%%%

\begin{abstract}
The spontaneous breaking of $U(1)_{B-L}$ 
around the scale of grand unification can simultaneously account for hybrid inflation, leptogenesis, and neutralino dark matter, thus resolving three major puzzles of particle physics and cosmology in a single predictive framework.
The $B\!-\!L$ phase transition also results in a network of cosmic strings.
If strong and electroweak interactions are unified in an $SO(10)$
gauge group, containing $U(1)_{B-L}$ as a subgroup, these strings are metastable.
In this case, they produce a stochastic background of gravitational waves that evades current pulsar timing bounds, but features a flat spectrum with amplitude $h^2\Omega_\text{GW} \sim 10^{-8}$ at interferometer frequencies.
Ongoing and future LIGO observations will hence probe the scale of $B\!-\!L$ breaking.
\end{abstract}

%%%%%%%%%%%%%%%%%%%%%%%%%%%%%%%%%%%%%%%%%%%%%%%%%%%%%%%%%%%%%%%%%%%%%%%%%%%%%%%%%%%%%%%%%%%%%%%%%%%%

\date{\today}
\maketitle

%%%%%%%%%%%%%%%%%%%%%%%%%%%%%%%%%%%%%%%%%%%%%%%%%%%%%%%%%%%%%%%%%%%%%%%%%%%%%%%%%%%%%%%%%%%%%%%%%%%%

\noindent\textbf{Introduction.}
The grand unified (GUT) gauge group $SO(10)$ contains $B\!-\!L$, the difference between baryon and lepton number, as a local symmetry.
As shown in Ref.~\cite{Buchmuller:2012wn}, the decay of a false vacuum of unbroken $B\!-\!L$ symmetry is an intriguing and testable mechanism to generate the initial conditions of the hot early universe (see Ref.~\cite{Buchmuller:2013dja} for a review).
With $B\!-\!L$ broken close to the unification scale,
the false-vacuum phase yields hybrid inflation~\cite{Linde:1993cn} and ends in tachyonic preheating~\cite{Felder:2000hj}.
Decays of the $B\!-\!L$ breaking Higgs field and thermal processes produce an abundance of heavy neutrinos whose decays generate the entropy of the hot early universe, the baryon asymmetry via leptogenesis~\cite{Fukugita:1986hr}, and dark matter (DM) in the form of the lightest supersymmetric particle (LSP)~\cite{Ellis:1983ew}.

%%%%%%%%%%%%%%%%%%%%%%%%%%%%%%%%%%%%%%%%%%%%%%%%%%%%%%%%%%%%%%%%%%%%%%%%%%%%%%%%%%%%%%%%%%%%%%%%%%%%

The cosmological evolution during and after inflation is determined by a small set of parameters, including the $B\!-\!L$ breaking scale and the masses of the $B\!-\!L$ Higgs boson, heavy neutrino, and gravitino.
In this Letter, we revisit the model in Ref.~\cite{Buchmuller:2012wn} and perform for the first time a global analysis identifying the viable parameter space that simultaneously explains the primordial power spectrum of the cosmic microwave background (CMB), the matter-antimatter asymmetry, and the DM relic density.

In addition, we study the network of cosmic strings (CSs) that is produced during the $B\!-\!L$ phase transition.
We embed our model in an $SO(10)$ GUT, which renders the $B\!-\!L$
strings metastable due to the nonperturbative (and hence exponentially suppressed) production of $SO(10)$ monopole--antimonopole pairs.
As a consequence, we find that the cosmic-string network generates a stochastic background of gravitational waves (GWs) that evades current pulsar timing constraints but that is testable in ongoing and future GW observations.
For a mild hierarchy between the $SO(10)$ and $U(1)_{B-L}$ breaking scales, LIGO can probe the scale of $B\!-\!L$ breaking.
The latter result is a direct consequence of the symmetry breaking pattern and is independent of many details of the model, resulting in a prediction of the GW spectrum in terms of the two dimensionful parameters of the cosmic string network. This represents a clear target for current and future GW experiments.

%%%%%%%%%%%%%%%%%%%%%%%%%%%%%%%%%%%%%%%%%%%%%%%%%%%%%%%%%%%%%%%%%%%%%%%%%%%%%%%%%%%%%%%%%%%%%%%%%%%%

\medskip\noindent\textbf{Supersymmetric \boldmath{$B\!-\!L$} model.} The example analyzed in Ref.~\cite{Buchmuller:2012wn} is based on the gauge group $SU(3)\times SU(2)\times U(1)_Y \times U(1)_{B-L} \equiv G_{\rm SM} \times U(1)_{B-L}$ and superpotential
\begin{align}
\label{W}
W & = W_{\rm MSSM} + h_{ij}^\nu \mathbf{5}_i^* n_j^c H_u + \frac{1}{\sqrt{2}}\,h_i^n n_i^c n_i^c S_1 
\nonumber \\ 
& + \lambda\,\Phi\left(\frac{v_{B-L}^2}{2} - S_1 S_2\right) + W_0 \,.
\end{align}
Here, $S_{1,2}$ are chiral superfields whose vacuum expectation values break $B\!-\!L$.
In unitary gauge, they correspond to the physical $B\!-\!L$ Higgs superfield, $S_{1,2} = S/\sqrt{2}$.
$\Phi$ is the inflaton superfield, $n^c_i$ contain the charge conjugates of the right-handed neutrinos, the Standard Model (SM) leptons are arranged in the $SU(5)$ multiplets $\mathbf{5}^* = (d^c,\ell)$ and $\mathbf{10} = (q,u^c,e^c)$, and the SM Higgs fields are contained in $H_{u,d}$.
$W_\text{MSSM}$ is the MSSM superpotential,
\begin{align}
W_\text{MSSM} = h^u_{ij} \mathbf{10}_i\mathbf{10}_j H_u + h^d_{ij}\mathbf{5}^*_i\mathbf{10}_j H_d \,,
\end{align}
while $W_0$ is a constant ensuring the correct zero point of the scalar potential in supergravity.

%%%%%%%%%%%%%%%%%%%%%%%%%%%%%%%%%%%%%%%%%%%%%%%%%%%%%%%%%%%%%%%%%%%%%%%%%%%%%%%%%%%%%%%%%%%%%%%%%%%%

The Yukawa couplings are chosen according to the Froggatt--Nielsen (FN) model~\cite{Froggatt:1978nt,Buchmuller:1998zf}, which is known to yield a satisfactory description of quark, charged-lepton, and neutrino masses and mixing angles, with
\begin{align}
h_{ij} \sim \eta^{Q_i+Q_j}\ , \quad \lambda \sim \eta^{Q_\Phi}\ ,
\end{align}
$\eta \simeq 1/\sqrt{300}$ and FN charges listed in Table~\ref{tab:FN}.
As usual, the Yukawa couplings have unspecified $\mathcal{O}(1)$ coefficients that are not $SU(5)$-symmetric.
For simplicity, the FN charges are restricted to $b = c = d-1 = e/2$.
This implies the following mass relations for the heavy Majorana neutrinos $N_{1,2,3}$ and the $B\!-\!L$ Higgs field $S$,
\begin{equation}
\label{massrelations}
\begin{split}
M_2 &\simeq M_3 \simeq m_S \simeq \eta^{-2} M_1\ , \\ M_1 & \sim \eta^{2d} \, v_{B-L}\ , \quad v_{B-L} \sim \eta^{2a} \frac{v^2_{\rm EW}}{\overline{m}_\nu} \ ,
\end{split}
\end{equation}
where $v_{B-L}$ is the $B\!-\!L$ breaking scale, $v_{\rm EW} \simeq 174\,\textrm{GeV}$ is the electroweak scale, and $\overline{m}_\nu = \sqrt{m_2m_3}$.
The decays $S \rightarrow N_2 N_2, N_3 N_3$ are thus forbidden, so that leptogenesis is dominated by the decay chain $S\rightarrow N_1 N_1, N_1 \rightarrow \ell H_u$.
The light neutrino masses $m_{1,2,3}$ (with normal hierarchy) are the singular values of the seesaw mass matrix,
\begin{align}
m_\nu = - m_D\,M^{-1}m_D^T\ ,
\end{align}
where $m_D$ and $M$ denote the Dirac and heavy Majorana neutrino mass matrices, respectively.
As we will discuss in more detail below, successful inflation requires $v_{B-L} \simeq \textrm{few} \times 10^{15}~\text{GeV}$, corresponding to $a = 0$.
Hence, top and bottom Yukawa couplings are of the same order in $\eta$, 
and $v_{\text{EW}} \simeq \left\langle H_u\right\rangle$.
Finally, an important quantity for leptogenesis is the effective light neutrino mass
\begin{align}
\widetilde{m}_1 = \frac{(m_D^\dagger m_D^{\vphantom{dagger}})_{11}}{M_1}\ .
\end{align}
Parametrically, one has $\widetilde{m}_1 \sim \overline{m}_\nu \simeq 0.02~\text{eV}$.
However, since $\widetilde{m}_1$ strongly depends on unknown $\mathcal{O}(1)$ factors for the Yukawa couplings, $\widetilde{m}_1$ is treated as a free parameter in the range $10^{-5}~\text{eV} \leq \widetilde{m}_1 \leq 10^{-1}~\text{eV}$.
In summary, the free parameters of the model can be chosen to be $v_{B-L}$, $M_1$, $\widetilde{m}_1$ and $W_0$.

%%%%%%%%%%%%%%%%%%%%%%%%%%%%%%%%%%%%%%%%%%%%%%%%%%%%%%%%%%%%%%%%%%%%%%%%%%%%%%%%%%%%%%%%%%%%%%%%%%%%

\begin{table}
\renewcommand{\arraystretch}{1.4}
\begin{center}
\caption{FN flavor charge assignment (from Ref.~\cite{Buchmuller:2012wn}).}
\label{tab:FN}
\begin{tabular}{c|cccccccccccc}
$\psi_i$  & $\mathbf{10}_3$  & $\mathbf{10}_2$ & $\mathbf{10}_1$ & $\mathbf{5}^*_3$ & $\mathbf{5}^*_2$ & $\mathbf{5}^*_1$ & $n^c_3$ & $n^c_2$ & $n^c_1$ & $H_{u,d}$ & $S_{1,2}$ & $\Phi$ \\\hline
$Q_i$     & 0 & 1 & 2 & $a$ & $a$ & $a+1$ & $b$ & $c$ & $d$ & 0 & 0 & $e$ 
\end{tabular}
\end{center}
\end{table}

%%%%%%%%%%%%%%%%%%%%%%%%%%%%%%%%%%%%%%%%%%%%%%%%%%%%%%%%%%%%%%%%%%%%%%%%%%%%%%%%%%%%%%%%%%%%%%%%%%%%

\medskip\noindent\textbf{GUT embedding and cosmic strings.} The embedding of $G_{\rm SM} \times U(1)_{B-L}$ into a larger GUT group, spontaneously broken at $v_\text{GUT} > v_{B-L}$, determines the  $B\!-\!L$ charges of $S_{1,2}$.
Ref.~\cite{Buchmuller:2012wn} is based on an $SU(5)$ FN model, and the $B\!-\!L$ charges are $q_S \equiv q_{S_2} = - q_{S_1} = 2$, $q_{n^c_i} =1$, $q_\Phi = 0$. In this case, the final unbroken gauge group is $G_{\rm SM} \times \mathbb{Z}_2$, which results in stable CSs~\cite{Dror:2019syi}. 
However, the model can be modified such that $q_S  = 1$.
This is achieved by the replacement in Eq.~\eqref{W}
\begin{align}\label{newmodel}
\frac{1}{\sqrt{2}} h_i^n n_i^c n_i^c S_1 \rightarrow \frac{1}{M_*} h_i^n n_i^c n_i^c S_1S_1 \,,
\end{align}
where the mass scale $M_*$ is larger than the GUT scale.
The heavy neutrino masses are now $M_i = h^n_i v_{B-L}^2/M_*$.
For $M_*$ close to $v_{B-L}$, the mass relations in Eq.~\eqref{massrelations} can be preserved by changing the $\mathcal{O}(1)$ factors.
In this form, the model can be embedded into an $SO(10)$ FN model~\cite{Asaka:2003fp}.
$S_1$ and $S_2$ become part of a $\mathbf{16}^*$ and $\mathbf{16}$, respectively, and a further $\mathbf{16}^* + \mathbf{16}$ pair together with a $\mathbf{10}$ is introduced to achieve an embedding of quarks and leptons that only respects $SU(5)$ but not $SO(10)$~\cite{Nomura:1998gm}.
In this way, the $SU(5)$ FN charges in Table~\ref{tab:FN} can be obtained~\cite{Asaka:2003fp}.

%%%%%%%%%%%%%%%%%%%%%%%%%%%%%%%%%%%%%%%%%%%%%%%%%%%%%%%%%%%%%%%%%%%%%%%%%%%%%%%%%%%%%%%%%%%%%%%%%%%%

With $q_S = 1$, the final unbroken group is $G_{\rm SM}$, and there can be no stable strings since the first homotopy group vanishes, $\Pi_1\left(SO(10)/G_{\rm SM}\right) = 0$~\cite{Dror:2019syi}.
To understand this, consider the breaking scheme (see also~\cite{Leblond:2009fq})
\begin{equation}
SO(10) \rightarrow G_\text{SM} \times U(1)_{B-L} \rightarrow G_\text{SM} \ .
\end{equation}
The first step produces monopoles, while the second step produces CSs.
The result is a network of CSs and monopoles, which is unstable~\cite{Vilenkin:1982hm, Martin:1996ea, Martin:1996cp}.
The more interesting scenario is if cosmic inflation occurs after GUT but before $B\!-\!L$ breaking.
In this case, the initial monopole population is diluted away.
The CSs can then only decay via the Schwinger production of monopole--antimonopole pairs, leading to a metastable CS network.
In this case, the decay rate per string unit length is~\cite{Leblond:2009fq, Monin:2008mp,Monin:2009ch}
\begin{equation}
\Gamma_d = \frac{\mu}{2 \pi} \exp\left( - \pi \kappa \right) \,,
\end{equation}
with $\kappa = m^2/\mu$ denoting the ratio between the monopole mass $m \sim v_\text{GUT}$ and the CS tension $\mu$.
In Abelian field theories, one obtains $\mu = 2 \pi B(\beta) v^2_{B-L}$ with $\beta = \lambda/(8 g^2)$ and $B(\beta) = 2.4/\ln(2/\beta)$ for $\beta < 0.01$~\cite{Hindmarsh:2011qj}.
For an appropriate choice of $v_{B-L} < v_{\rm GUT}$, the CSs are sufficiently long-lived to give interesting signatures but decay before emitting low-frequency GWs that are strongly constrained by pulsar timing arrays.

%%%%%%%%%%%%%%%%%%%%%%%%%%%%%%%%%%%%%%%%%%%%%%%%%%%%%%%%%%%%%%%%%%%%%%%%%%%%%%%%%%%%%%%%%%%%%%%%%%%%

\medskip\noindent\textbf{Hybrid inflation, baryogenesis, dark matter.}
The superpotential term in Eq.~\eqref{W} causing spontaneous $B\!-\!L$ breaking is precisely the superpotential of F-term hybrid inflation (FHI)~\cite{Copeland:1994vg,Dvali:1994ms}.
It was widely believed that FHI could not account for the correct scalar spectral index of the CMB power spectrum; but the analyses in
\cite{BasteroGil:2006cm,Rehman:2009nq,Nakayama:2010xf,Buchmuller:2014epa,Schmitz:2018nhb} showed that FHI is viable once the effect of supersymmetry (SUSY) breaking on the inflaton potential is taken into account. 
With an (almost) vanishing cosmological constant, SUSY breaking generates the constant term
\begin{align}\label{W0}
W_0 = \alpha\,m_{3/2}\,M_{\rm Pl} \,,
\end{align}
where $\alpha \sim 1$ encodes the details of SUSY breaking and $m_{3/2}$ is the gravitino mass.
Choosing $\alpha=1$ for definiteness, this adds
%(the sum of a constant, a radiative Coleman-Weinberg contribution and a supergravity part)
a term linear in the inflaton field to the inflaton potential~\cite{Buchmuller:2000zm},
\begin{align}
V_{3/2} (\Phi) = - \lambda\,v_{B-L}^2\,m_{3/2}\left(\Phi + \Phi^*\right) + \cdots \,.
\end{align}
FHI now becomes a two-field model of inflation in the complex $\Phi$-plane~\cite{Buchmuller:2014epa}.
The choice of the inflationary trajectory impacts the CMB observables and even determines whether a graceful exit from inflation is at all possible.
Trajectories far from the real axis require a significant amount of tuning.
For simplicity, we thus restrict ourselves to trajectories along the real axis.
In this case, the observed values of the amplitude and index of the scalar power spectrum, $A_s^{\rm obs}$ and $n_s^{\rm obs}$, eliminate two of the three parameters $v_{B-L}$, $\lambda$, and $m_{3/2}$, where we recall that we can exchange $\lambda \leftrightarrow M_1$ using the FN relations~\eqref{massrelations}.

%%%%%%%%%%%%%%%%%%%%%%%%%%%%%%%%%%%%%%%%%%%%%%%%%%%%%%%%%%%%%%%%%%%%%%%%%%%%%%%%%%%%%%%%%%%%%%%%%%%%

We will now demonstrate that the parameter values required for FHI are consistent with leptogenesis and neutralino DM.
First, we define the reheating temperature $T_{\text{rh}}$ in terms of the total and radiative energy densities,
\begin{align}\label{Trh}
T = T_\text{rh} \quad\Leftrightarrow\quad \rho_\text{tot} = 2\,\rho_\text{rad} \,.
\end{align}
$T_\text{rh}$ is determined by the decay widths of $S$ and $N_1$, which can be expressed in terms of the masses defined above,
\begin{align}
 \Gamma^0_S & = \frac{m_S}{8\pi}\left(\frac{M_1}{v_{B-L}}\right)^2\left(1-\frac{4 M^2_1}{m^2_S}\right)^{1/2} \,,
\\ \nonumber
\Gamma^0_{N_1} & \simeq \frac{1}{4\pi}\frac{\widetilde{m}_1}{\overline{m}_\nu}\frac{M_1^2}{v_{B-L}} \,.
\end{align}
The neutrinos $N_1$ produced in $S$ decays are relativistic, which is accounted for by the averaged Lorentz factor $\gamma$,
\begin{align}
\Gamma^S_{N_1} = \gamma^{-1}\,\Gamma^0_{N_1}\ , \quad  \gamma^{-1} = \left\langle \frac{M_1}{E_{N_1}} \right\rangle\ .
\end{align}
%
%with $E_{N_1}$ denoting the $N_1$ energy.
%
Equating the decay rates
$\Gamma_S \equiv \Gamma^0_S$ and $\Gamma_{N_1} \equiv  \Gamma^S_{N_1}$ with the Hubble parameter $H$ defines the $S$ and $N_1$ decay temperatures,
\begin{align}
H\left(T_X\right) = \Gamma_X  :\:\:
T_X = \left(\frac{90}{\alpha_X\pi^2g_*}\right)^{1/4}\sqrt{\Gamma_X M_{\rm Pl}} \,,
\end{align}
where $X = \{S,N_1\}$ and $\alpha^{-1}_X \equiv \rho_{\text{rad}}/\rho_{\text{tot}}|_{T_X}$. 
For $T_{\rm rh}$, we then find (see Appendix~C in~\cite{Buchmuller:2013lra} for details)
\begin{align}
T_{\mathrm{rh}} &\simeq 0.85 \times \mathrm{min}\{T_{N_1},T_S\} \,.
\end{align}

%%%%%%%%%%%%%%%%%%%%%%%%%%%%%%%%%%%%%%%%%%%%%%%%%%%%%%%%%%%%%%%%%%%%%%%%%%%%%%%%%%%%%%%%%%%%%%%%%%%%

\begin{figure}
\begin{center}
\includegraphics[width=0.48\textwidth]{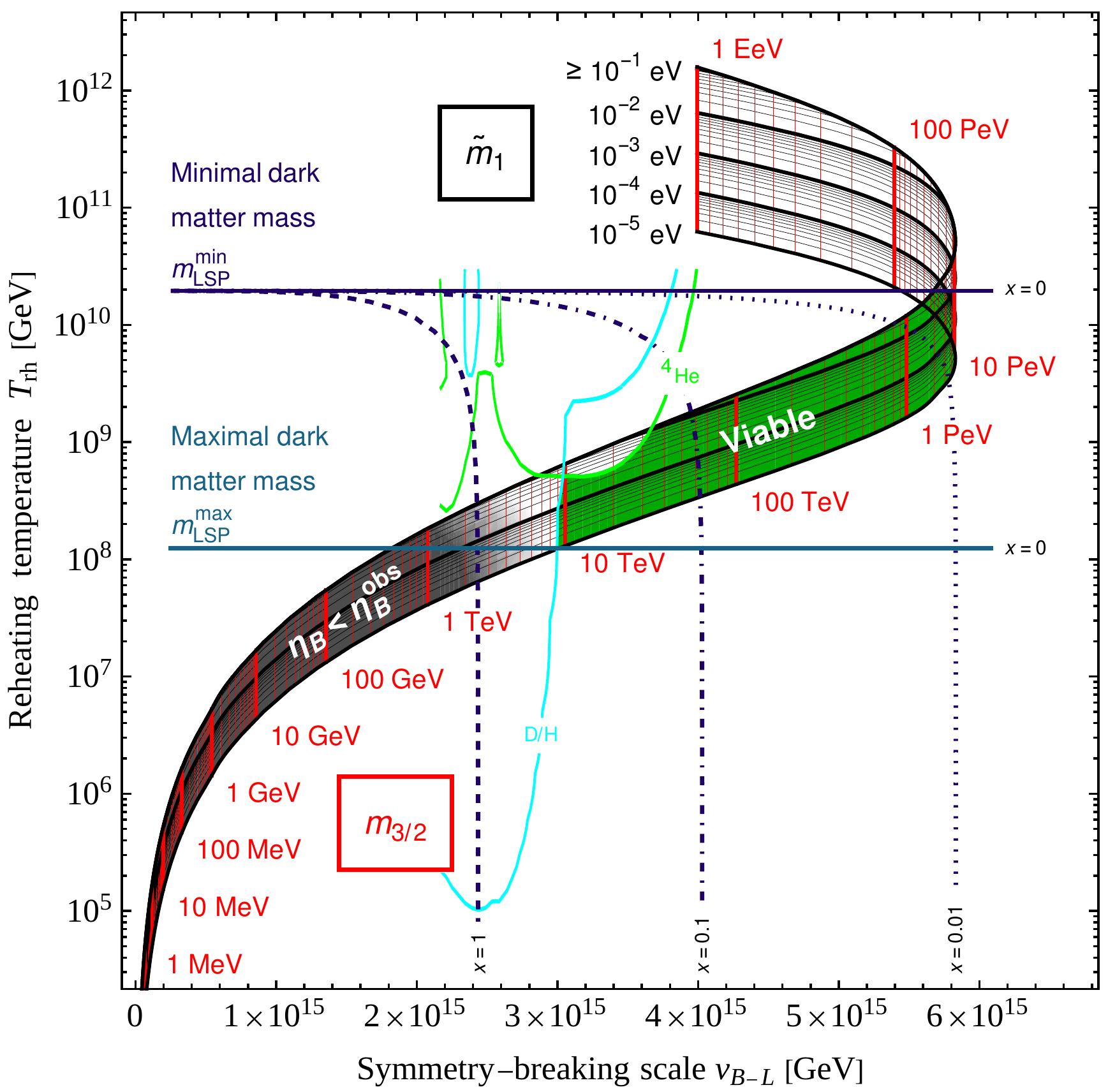}
\caption{Viable parameter space (green) for hybrid inflation, leptogenesis, neutralino DM, and big bang nucleosynthesis.
Hybrid inflation and the dynamics of reheating correlate the parameters $v_{B-L}$, $T_\text{rh}$, $m_{3/2}$ and $\widetilde{m}_1$ (black curves).
Successful leptogenesis occurs outside the gray-shaded region.
Neutralino DM is viable in the green region, corresponding to a higgsino (wino) with mass $100 \leq m_{\rm LSP}/\textrm{GeV} \leq 1060$ (2680).}
\label{fig:overlay}
\end{center}
\end{figure}

%%%%%%%%%%%%%%%%%%%%%%%%%%%%%%%%%%%%%%%%%%%%%%%%%%%%%%%%%%%%%%%%%%%%%%%%%%%%%%%%%%%%%%%%%%%%%%%%%%%%

For any given values of $\widetilde{m}_1$ and $m_{3/2}$, successful hybrid inflation selects a point in the $v_{B-L}$\,--\,$T_\text{rh}$ plane, such that the values of all parameters are fixed [up to an $\mathcal{O}(1)$ uncertainty due to the constant $\alpha$ in Eq.~\eqref{W0}].
This is illustrated in Figure~\ref{fig:overlay}. 
The turn-over at large gravitino masses and high reheating temperatures reflects the quartic supergravity coupling in the inflaton potential
becoming important.
Even larger gravitino masses entail field excursions of ${\cal O}(M_\text{Pl})$ and thus are sensitive to further, model-dependent supergravity corrections.
For large values of $\widetilde{m}_1$, the reheating temperature is set by $S$ decays, such that it becomes independent of $\widetilde{m}_1$.
This is reflected in the merging of different $\widetilde{m}_1$ contour lines at the upper edge of the band depicted in Fig.~\ref{fig:overlay}.

%%%%%%%%%%%%%%%%%%%%%%%%%%%%%%%%%%%%%%%%%%%%%%%%%%%%%%%%%%%%%%%%%%%%%%%%%%%%%%%%%%%%%%%%%%%%%%%%%%%%

The decay of the heavy neutrino $N_1$ is responsible for nonthermal leptogenesis~\cite{Lazarides:1991wu,Asaka:1999yd},
which dominates in our model over thermal leptogenesis for $T_\text{rh} \lesssim 10^{10}$~GeV~\cite{Buchmuller:2012wn}.
The resulting baryon asymmetry can be estimated as
\begin{align}
\label{eq:etaB}
\eta_B \simeq
\eta_B^{\rm nt} \simeq C_{\rm sph}\,\frac{g_{*,s}^0}{g_{*,s}}\frac{g_{*,\rho}}{g_\gamma}\frac{\pi^4}{30\,\zeta(3)}\,\varepsilon_1 \frac{T_{\rm rh}}{\gamma\,M_1} \,,
\end{align}
with $C_\text{sph} = 8/23$ denoting the sphaleron conversion factor, $g_{*,\rho} = g_{*,s} = 915/4$, $g_{*,s}^0 = 43/11$, and $g_\gamma = 2$ counting (effective) numbers of relativistic degrees of freedom and $\varepsilon_1 \lesssim 2 \times 10^{-6} M_1/\left(10^{10}\text{GeV}\right)$~\cite{Hamaguchi:2001gw,Davidson:2002qv,Buchmuller:2005eh} parameterizing the $CP$ asymmetry in $N_1$ decays.
This agrees with the result obtained solving the corresponding Boltzmann equations within a factor of two~\cite{Buchmuller:2012wn}.
The gray shading in Fig.~\ref{fig:overlay} indicates the region where leptogenesis falls short of explaining the observed baryon asymmetry.
%

%%%%%%%%%%%%%%%%%%%%%%%%%%%%%%%%%%%%%%%%%%%%%%%%%%%%%%%%%%%%%%%%%%%%%%%%%%%%%%%%%%%%%%%%%%%%%%%%%%%%

Gravitino masses of $\mathcal{O}\left(1\right)\,\textrm{TeV}$ or larger point to a neutralino LSP, which is produced thermally as well as nonthermally in gravitino decays~\cite{Buchmuller:2012bt}.
Gravitinos are in turn generated in decays of the $B\!-\!L$ Higgs field at a rate
\begin{equation}
\label{eq:Gamma32}
\Gamma_S^{3/2} = \frac{x}{32 \pi} \left(\frac{\left<S\right>}{M_\text{Pl}}\right)^2 \frac{m_S^3}{M_\text{Pl}^2} \,,
\end{equation}
as well as from the thermal bath (for a discussion and references, see~\cite{Jeong:2012en}).
The parameter $x$ in Eq.~\eqref{eq:Gamma32} encodes details of the unspecified SUSY-breaking sector, and for definiteness, we will assume all gaugino masses to be significantly lighter than $m_{3/2}$.
Taking into account that gravitinos must decay early enough to preserve big bang nucleosynthesis (BBN)~\cite{Kawasaki:2017bqm} and $m_\text{LSP} \gtrsim 100$~GeV~\cite{PhysRevD.98.030001}, a higgsino or wino LSP can account for the observed DM relic density in the green-shaded region of Fig.~\ref{fig:overlay}. 
It is highly nontrivial that neutralino DM and leptogenesis can be successfully realized in the same parameter region.
In summary, the viable parameter region of our model is given by $v_{B-L} \simeq 3.0\cdots5.8\times10^{15}\,\textrm{GeV}$, $m_{3/2} \simeq 10\,\textrm{TeV}\cdots10\,\textrm{PeV}$, and $x \lesssim 0.4$.

%%%%%%%%%%%%%%%%%%%%%%%%%%%%%%%%%%%%%%%%%%%%%%%%%%%%%%%%%%%%%%%%%%%%%%%%%%%%%%%%%%%%%%%%%%%%%%%%%%%%

\medskip\noindent\textbf{Gravitational waves.}
The network of CSs formed during the $B\!-\!L$ phase transition acts
as a source of GWs~\cite{Buchmuller:2013lra}.
The dominant contribution is expected to come from long-lived, sub-horizon CS loops that eventually decay emitting gravitational radiation.
Modeling the evolution and GW emission of a CS network is a challenging task, resulting in several competing models in the literature.
Among others, major questions are the ansatz for CSs (field theory versus Nambu--Goto), the correct loop number density and the most efficient GW production channel (see Ref.~\cite{Auclair:2019wcv} and references therein for a comprehensive review).
Moreover, for metastable strings,
the GW production from fast-moving monopoles requires further investigation~\cite{Vilenkin:1982hm,Leblond:2009fq}.
For concreteness, we will base our analysis on the BOS model~\cite{Blanco-Pillado:2013qja}, but the arguments presented here can easily be applied to any other CS model.
%

%%%%%%%%%%%%%%%%%%%%%%%%%%%%%%%%%%%%%%%%%%%%%%%%%%%%%%%%%%%%%%%%%%%%%%%%%%%%%%%%%%%%%%%%%%%%%%%%%%%%

\begin{figure}
\begin{center}
\includegraphics[width=0.48\textwidth]{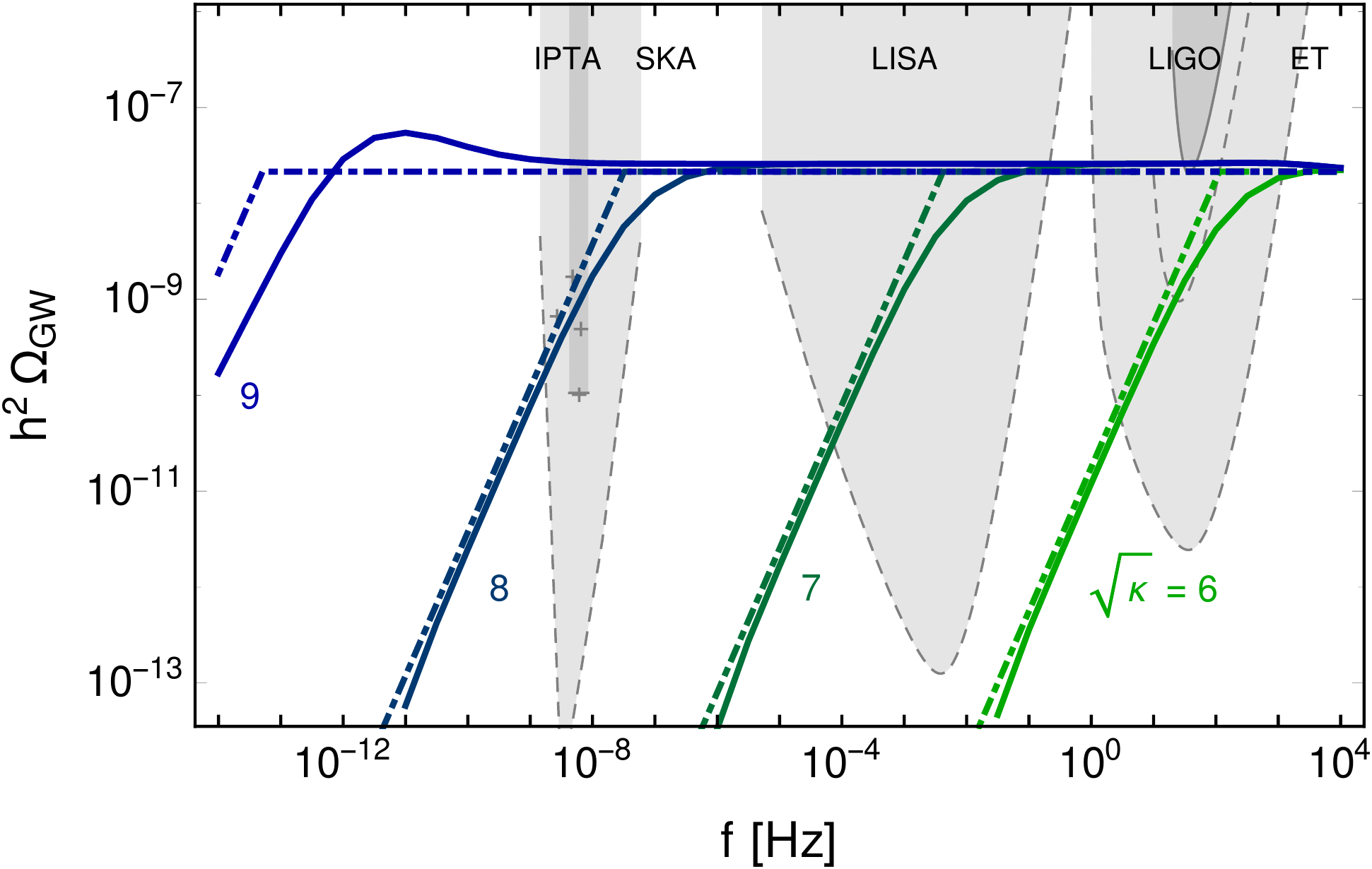}
\caption{GW spectrum for $G \mu = 2 \times 10^{-7}$.
Different values of $\sqrt{\kappa}$ are indicated in different colors; the blue curve corresponds to a CS network surviving until today.
The dot-dashed lines depict the analytical estimate~\eqref{eq:Omegaf_ana}.
The (lighter) gray-shaded areas indicate the sensitivities of (planned) experiments SKA~\cite{Smits:2008cf}, LISA~\cite{Audley:2017drz}, LIGO~\cite{LIGOScientific:2019vic} and ET~\cite{ET}, the crosses within the SKA band indicate constraints by the IPTA~\cite{Verbiest:2016vem}.}
\label{fig:GWspectrum}
\end{center}
\end{figure}

%%%%%%%%%%%%%%%%%%%%%%%%%%%%%%%%%%%%%%%%%%%%%%%%%%%%%%%%%%%%%%%%%%%%%%%%%%%%%%%%%%%%%%%%%%%%%%%%%%%%

The present-day GW spectrum can be expressed as~\cite{Auclair:2019wcv}
\begin{align}
\Omega_\text{GW}(f) = \frac{\partial \rho_\text{GW}(f)}{\rho_c \partial \ln f}= \frac{8 \pi f (G \mu)^2}{3 H_0^2} \sum_{n = 1}^\infty C_n(f) \, P_n \,,
\label{eq:Omega}
\end{align}
where $\rho_\text{GW}$ denotes the GW energy density, $\rho_c$ is the critical energy density of the universe, $G \mu$ denotes the dimensionless string tension, $H_0 = 100 \,h\,\textrm{km}/\textrm{s}/\textrm{Mpc}$ is today's Hubble parameter, $P_n \simeq 50/\zeta[4/3] \,n^{-4/3}$ is the power spectrum of GWs emitted by the $n^{\rm th}$ harmonic of a CS loop, and $C_n(f)$ indicates the number of loops emitting GWs that are observed at a given frequency $f$,
\begin{align}
\label{eq:Cn}
C_n(f) = \frac{2 n}{f^2} \int_{z_\text{min}}^{z_\text{max}}dz\:\frac{\mathcal{N}\left(\ell\left(z\right),\,t\left(z\right)\right)}{H\left(z\right)(1 + z)^6} \,,
\end{align}
which is a function of the number density of CS loops $\mathcal{N}(\ell,t)$, with $\ell = 2n/((1 + z) f)$, selecting the loops that contribute to the spectrum at frequency $f$ today.
The loop number density can be estimated analytically and improved with input from numerical simulations.
In particular, for loops generated and decaying in the radiation-dominated (RD) era, it can be estimated as~\cite{Blanco-Pillado:2013qja,Auclair:2019wcv}
\begin{align}
\mathcal{N}_r(\ell,t) & = \frac{0.18}{t^{3/2} (\ell + \Gamma G \mu t)^{5/2}}  \,, \label{eq:nr}
\end{align}
where $\Gamma \simeq 50$ parametrizes the CS decay rate into GWs, $\dot \ell = - \Gamma G \mu$.
The integration range in Eq.~\eqref{eq:Cn} accounts for the lifetime of the CS network, from the formation at $z_\text{max} \simeq T_\text{rh}/(2.7~\text{K})$ until their decay when the decay rate of a string loop with average length $\bar \ell \simeq 13 \, G \mu/H$ equals the Hubble rate, $\bar \ell  \, \Gamma_d = H$. The latter condition yields
\begin{equation}
z_\text{min} = \left( \frac{70}{H_0}\right)^{1/2} \left( \Gamma  \; \Gamma_d  \; G \mu \right)^{1/4} \,,  
\label{eq:zmin}
\end{equation}
which coincides with the estimate found in~\cite{Leblond:2009fq}. 

%%%%%%%%%%%%%%%%%%%%%%%%%%%%%%%%%%%%%%%%%%%%%%%%%%%%%%%%%%%%%%%%%%%%%%%%%%%%%%%%%%%%%%%%%%%%%%%%%%%%

\begin{figure}
\centering
\includegraphics[width = 0.48\textwidth]{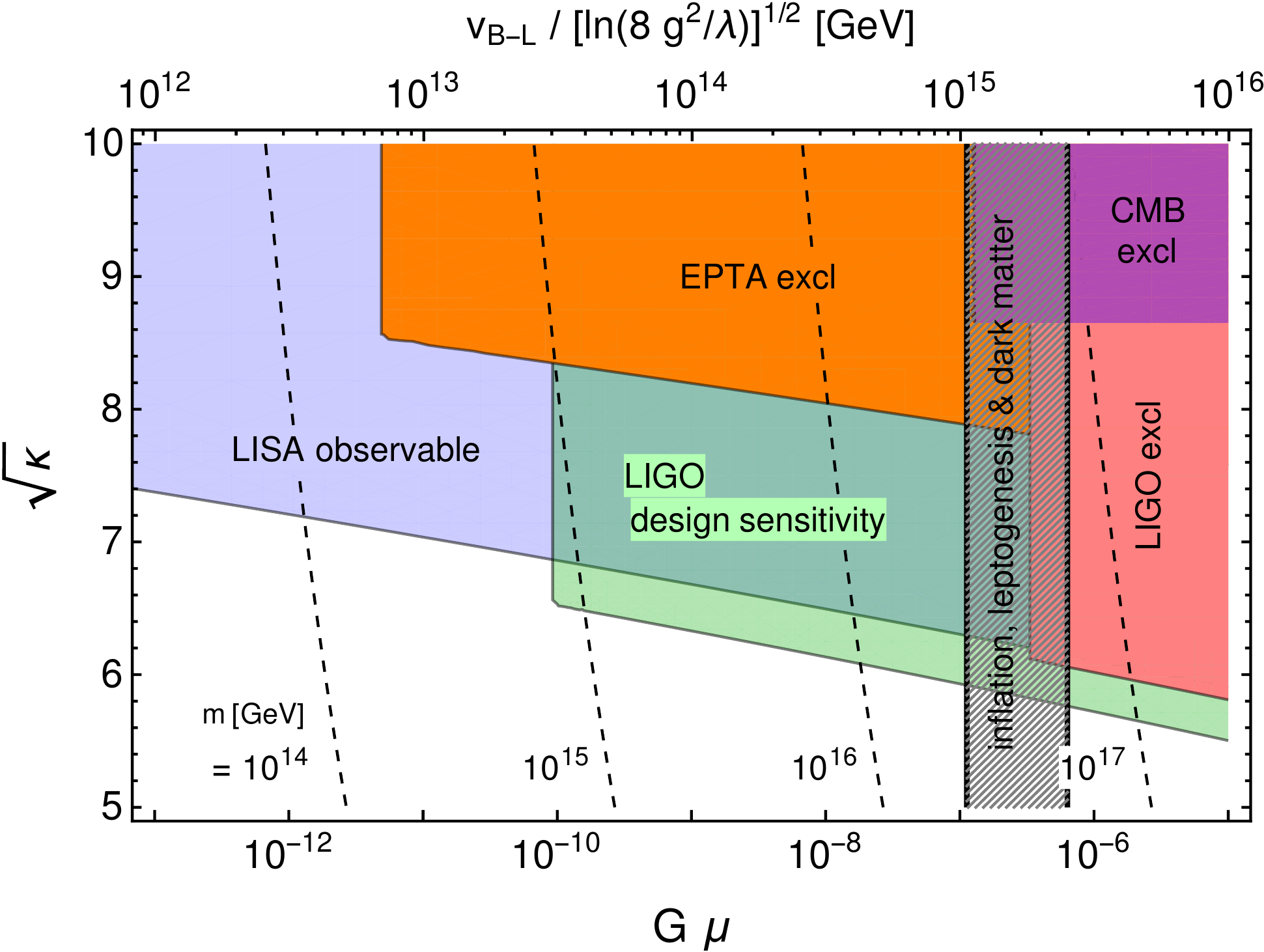}
\caption{Constraints on the CS parameter space from GW searches.
The orange, red, and purple regions are excluded by the existing bounds from EPTA~\cite{Shannon:2015ect}, LIGO O2~\cite{LIGOScientific:2019vic} and PLANCK~\cite{Ade:2015xua}, respectively.
The shaded green and blue regions indicate the prospective reach of
LIGO at design sensitivity and of LISA. The hatched region indicates
the viable parameter space from Fig.~\ref{fig:overlay}.}
\label{fig:CSconstraints}
\end{figure}

%%%%%%%%%%%%%%%%%%%%%%%%%%%%%%%%%%%%%%%%%%%%%%%%%%%%%%%%%%%%%%%%%%%%%%%%%%%%%%%%%%%%%%%%%%%%%%%%%%%%

For a sufficiently long-lived CS network, the spectrum features a plateau over many orders of magnitude in frequency, accounting for the GW production from CS loops during the RD era.
The amplitude of this plateau is~\cite{Auclair:2019wcv}
\begin{equation}
\label{eq:Omega_ana}
\Omega_\text{GW}^\text{plateau} \simeq 8.04 \, \Omega_r \left( \frac{G \mu}{\Gamma} \right)^{1/2} \,.
\end{equation}
A finite lifetime of the CS network leads to a suppression of the GW spectrum at low frequencies.
Focusing for simplicity only on the RD era, we note that the scaling of the integrand of Eq.~\eqref{eq:Cn} with $z$ changes at $z_\ell = \Gamma/2\,G \mu\,f\,t_\text{ref}\, z_\text{ref}^2$, with $t_\text{ref} = t(z_\text{ref})$ denoting an arbitrary reference time during RD.
One can verify that, for $z_\text{min} \ll z_{\ell}$ (corresponding to large frequencies), $\Omega_\text{GW}(f)$ becomes scale-invariant, whereas for $z_\text{min} \gg z_{\ell}$, we find $\Omega_\text{GW} \propto f^{3/2}$.
With this, we can estimate the turn-over point $f_*$ of the spectrum by solving $z_\text{min} = z_\ell$ for $f$,
\begin{align}
\label{eq:fs_ana}
f_* \simeq 3.0 \times 10^{14} \, \text{Hz} \:e^{- \pi \kappa/4} \left(\frac{ 10^{-7}}{G \mu }\right)^{1/2} \,.
\end{align}
%
%It is worth noting that, for long (i.e., horizon-sized) CSs, the corresponding turn-over frequency is much lower, $f_*^\text{LS} \simeq 10^9~\text{Hz } \left(G\mu / 10^{-7}\right)^{1/4} e^{-\pi\kappa/4}$.
%
%In principle, this could enable one to probe this irreducible but subdominant part of the GW spectrum at low frequencies.
%
In summary, we arrive at a simple estimate for the GW spectrum from
a metastable CS network,
\begin{align}
\label{eq:Omegaf_ana}
\Omega_\text{GW}(f) = \Omega_\text{GW}^\text{plateau} \; \text{min}\left[ (f/f_*)^{3/2},1 \right] \,.
\end{align}
Fig.~\ref{fig:GWspectrum} shows the GW spectrum obtained by numerically evaluating Eq.~\eqref{eq:Omega} (see Ref.~\cite{Auclair:2019wcv} for details) as well as the analytical estimate Eq.~\eqref{eq:Omegaf_ana}.
The shaded regions indicate the power-law-integrated sensitivity curves of current and planned experiments~\cite{Thrane:2013oya}.
We see that, for $G \mu = 2 \times 10^{-7}$, the constraint from the European Pulsar Timing Array (EPTA)~\cite{Shannon:2015ect} enforces $\sqrt{\kappa} \lesssim 8$.
The analytical estimate accurately explains the plateau value and the $f^{3/2}$ drop-off at low frequencies, showing overall good agreement with the numerical result.
%

%%%%%%%%%%%%%%%%%%%%%%%%%%%%%%%%%%%%%%%%%%%%%%%%%%%%%%%%%%%%%%%%%%%%%%%%%%%%%%%%%%%%%%%%%%%%%%%%%%%%

Comparing the predicted GW spectrum~\eqref{eq:Omegaf_ana} with the observational bounds and prospects depicted in Fig.~\ref{fig:GWspectrum}, we can map out the regions in the $\kappa$\,--\,$G \mu$ parameter plane that are already excluded or that will be probed in the near future, see Fig.~\ref{fig:CSconstraints}.
In particular, cosmological $B\!-\!L$ breaking with stable CSs is excluded, as are GUT monopole masses above $m = 5.4 \times 10^{16}$~GeV. Assuming a mild hierarchy between the GUT and $B\!-\!L$ scales, $m/v_{B-L} \gtrsim 6$, the entire remaining parameter space is testable with LIGO.

%%%%%%%%%%%%%%%%%%%%%%%%%%%%%%%%%%%%%%%%%%%%%%%%%%%%%%%%%%%%%%%%%%%%%%%%%%%%%%%%%%%%%%%%%%%%%%%%%%%%

\medskip\noindent\textbf{Conclusions.}
We have presented a minimal extension of the SM with $U(1)_{B-L}$
symmetry that
simultaneously explains inflation, leptogenesis, and neutralino DM.
Gravitinos are unstable and heavier than $10~\text{TeV}$.
Remarkably, the viable parameter space of our model automatically implies a large stochastic signal in GWs that is well within the reach of LIGO's design sensitivity.
It would be interesting to relax some of our model-building
assumptions, such as the underlying flavor model, in future work.
However, the general philosophy behind our model\,---\,inflation ending in a GUT-scale phase transition in combination with leptogenesis and dark matter in a SUSY extension of the SM\,---\,provides a testable framework for the physics of the early universe.
A characteristic feature of this framework is a stochastic background
of gravitational waves emitted by metastable cosmic strings.

%%%%%%%%%%%%%%%%%%%%%%%%%%%%%%%%%%%%%%%%%%%%%%%%%%%%%%%%%%%%%%%%%%%%%%%%%%%%%%%%%%%%%%%%%%%%%%%%%%%%

%\newpage
\vspace{0.1cm}
%\noindent 
\begin{acknowledgments}
\textit{Acknowledgements.} We thank Ryusuke Jinno for helpful discussions on the GW production from metastable cosmic strings.
This project has received funding from the Deutsche Forschungsgemeinschaft under Germany's Excellence Strategy\,--\,EXC 2121 ``Quantum Universe''\,--\,390833306 (V.\,D.) and
from the European Union's Horizon 2020 Research and Innovation Programme under grant agreement number 796961, ``AxiBAU'' (K.\,S.).
The work of H.\,M.\ was supported by the NSF grant PHY-1915314, by the U.S. DOE Contract DE-AC02-05CH11231, by the JSPS Grant-in-Aid for Scientific Research JP17K05409, MEXT Grant-in-Aid for Scientific Research on Innovative Areas JP15H05887, JP15K21733, by WPI, MEXT, Japan, and Hamamatsu Photonics.
\end{acknowledgments}

%%%%%%%%%%%%%%%%%%%%%%%%%%%%%%%%%%%%%%%%%%%%%%%%%%%%%%%%%%%%%%%%%%%%%%%%%%%%%%%%%%%%%%%%%%%%%%%%%%%%

\bibliographystyle{JHEP}
\bibliography{arxiv_2}

\providecommand{\href}[2]{#2}\begingroup\raggedright\begin{thebibliography}{10}

\bibitem{Buchmuller:2012wn}
W.~Buchmuller, V.~Domcke, and K.~Schmitz, {\it {Spontaneous $B\!-\!L$ Breaking
  as the Origin of the Hot Early Universe}},  {\em Nucl. Phys.} {\bf B862}
  (2012) 587--632, [\href{http://arxiv.org/abs/1202.6679}{{\tt
  arXiv:1202.6679}}].

\bibitem{Buchmuller:2013dja}
W.~Buchmuller, V.~Domcke, K.~Kamada, and K.~Schmitz, {\it {A Minimal
  Supersymmetric Model of Particle Physics and the Early Universe}},  in {\em
  {Cosmology and Particle Physics beyond Standard Models}}
  (L.~\'Avarez-Gaum\'e, G.~S. Djordjevi\'c, and D.~Stojkovi\'c, eds.),
  CERN-Proceedings-2014-001.
\newblock \href{http://arxiv.org/abs/1309.7788}{{\tt arXiv:1309.7788}}.

\bibitem{Linde:1993cn}
A.~D. Linde, {\it {Hybrid inflation}},  {\em Phys. Rev.} {\bf D49} (1994)
  748--754, [\href{http://arxiv.org/abs/astro-ph/9307002}{{\tt
  astro-ph/9307002}}].

\bibitem{Felder:2000hj}
G.~N. Felder, J.~Garcia-Bellido, P.~B. Greene, L.~Kofman, A.~D. Linde, and
  I.~Tkachev, {\it {Dynamics of symmetry breaking and tachyonic preheating}},
  {\em Phys. Rev. Lett.} {\bf 87} (2001) 011601,
  [\href{http://arxiv.org/abs/hep-ph/0012142}{{\tt hep-ph/0012142}}].

\bibitem{Fukugita:1986hr}
M.~Fukugita and T.~Yanagida, {\it {Baryogenesis Without Grand Unification}},
  {\em Phys. Lett.} {\bf B174} (1986) 45--47.

\bibitem{Ellis:1983ew}
J.~R. Ellis, J.~S. Hagelin, D.~V. Nanopoulos, K.~A. Olive, and M.~Srednicki,
  {\it {Supersymmetric Relics from the Big Bang}},  {\em Nucl. Phys.} {\bf
  B238} (1984) 453--476. [,223(1983)].

\bibitem{Froggatt:1978nt}
C.~D. Froggatt and H.~B. Nielsen, {\it {Hierarchy of Quark Masses, Cabibbo
  Angles and CP Violation}},  {\em Nucl. Phys.} {\bf B147} (1979) 277--298.

\bibitem{Buchmuller:1998zf}
W.~Buchmuller and T.~Yanagida, {\it {Quark lepton mass hierarchies and the
  baryon asymmetry}},  {\em Phys. Lett.} {\bf B445} (1999) 399--402,
  [\href{http://arxiv.org/abs/hep-ph/9810308}{{\tt hep-ph/9810308}}].

\bibitem{Dror:2019syi}
J.~A. Dror, T.~Hiramatsu, K.~Kohri, H.~Murayama, and G.~White, {\it {Testing
  the Seesaw Mechanism and Leptogenesis with Gravitational Waves}},  {\em Phys.
  Rev. Lett.} {\bf 124} (2020) 041804,
  [\href{http://arxiv.org/abs/1908.03227}{{\tt arXiv:1908.03227}}].

\bibitem{Asaka:2003fp}
T.~Asaka, {\it {Lopsided mass matrices and leptogenesis in SO(10) GUT}},  {\em
  Phys. Lett.} {\bf B562} (2003) 291--298,
  [\href{http://arxiv.org/abs/hep-ph/0304124}{{\tt hep-ph/0304124}}].

\bibitem{Nomura:1998gm}
Y.~Nomura and T.~Yanagida, {\it {Bimaximal neutrino mixing in SO(10)(GUT)}},
  {\em Phys. Rev.} {\bf D59} (1999) 017303,
  [\href{http://arxiv.org/abs/hep-ph/9807325}{{\tt hep-ph/9807325}}].

\bibitem{Leblond:2009fq}
L.~Leblond, B.~Shlaer, and X.~Siemens, {\it {Gravitational Waves from Broken
  Cosmic Strings: The Bursts and the Beads}},  {\em Phys. Rev.} {\bf D79}
  (2009) 123519, [\href{http://arxiv.org/abs/0903.4686}{{\tt
  arXiv:0903.4686}}].

\bibitem{Vilenkin:1982hm}
A.~Vilenkin, {\it {Cosmological evolution of monopoles connected by strings}},
  {\em Nucl. Phys.} {\bf B196} (1982) 240--258.

\bibitem{Martin:1996ea}
X.~Martin and A.~Vilenkin, {\it {Gravitational wave background from hybrid
  topological defects}},  {\em Phys. Rev. Lett.} {\bf 77} (1996) 2879--2882,
  [\href{http://arxiv.org/abs/astro-ph/9606022}{{\tt astro-ph/9606022}}].

\bibitem{Martin:1996cp}
X.~Martin and A.~Vilenkin, {\it {Gravitational radiation from monopoles
  connected by strings}},  {\em Phys. Rev.} {\bf D55} (1997) 6054--6060,
  [\href{http://arxiv.org/abs/gr-qc/9612008}{{\tt gr-qc/9612008}}].

\bibitem{Monin:2008mp}
A.~Monin and M.~B. Voloshin, {\it {The Spontaneous breaking of a metastable
  string}},  {\em Phys. Rev.} {\bf D78} (2008) 065048,
  [\href{http://arxiv.org/abs/0808.1693}{{\tt arXiv:0808.1693}}].

\bibitem{Monin:2009ch}
A.~Monin and M.~B. Voloshin, {\it {Destruction of a metastable string by
  particle collisions}},  {\em Phys. Atom. Nucl.} {\bf 73} (2010) 703--710,
  [\href{http://arxiv.org/abs/0902.0407}{{\tt arXiv:0902.0407}}].

\bibitem{Hindmarsh:2011qj}
M.~Hindmarsh, {\it {Signals of Inflationary Models with Cosmic Strings}},  {\em
  Prog. Theor. Phys. Suppl.} {\bf 190} (2011) 197--228,
  [\href{http://arxiv.org/abs/1106.0391}{{\tt arXiv:1106.0391}}].

\bibitem{Copeland:1994vg}
E.~J. Copeland, A.~R. Liddle, D.~H. Lyth, E.~D. Stewart, and D.~Wands, {\it
  {False vacuum inflation with Einstein gravity}},  {\em Phys. Rev.} {\bf D49}
  (1994) 6410--6433, [\href{http://arxiv.org/abs/astro-ph/9401011}{{\tt
  astro-ph/9401011}}].

\bibitem{Dvali:1994ms}
G.~R. Dvali, Q.~Shafi, and R.~K. Schaefer, {\it {Large scale structure and
  supersymmetric inflation without fine tuning}},  {\em Phys. Rev. Lett.} {\bf
  73} (1994) 1886--1889, [\href{http://arxiv.org/abs/hep-ph/9406319}{{\tt
  hep-ph/9406319}}].

\bibitem{BasteroGil:2006cm}
M.~Bastero-Gil, S.~F. King, and Q.~Shafi, {\it {Supersymmetric Hybrid Inflation
  with Non-Minimal Kahler potential}},  {\em Phys. Lett.} {\bf B651} (2007)
  345--351, [\href{http://arxiv.org/abs/hep-ph/0604198}{{\tt hep-ph/0604198}}].

\bibitem{Rehman:2009nq}
M.~U. Rehman, Q.~Shafi, and J.~R. Wickman, {\it {Supersymmetric Hybrid
  Inflation Redux}},  {\em Phys. Lett.} {\bf B683} (2010) 191--195,
  [\href{http://arxiv.org/abs/0908.3896}{{\tt arXiv:0908.3896}}].

\bibitem{Nakayama:2010xf}
K.~Nakayama, F.~Takahashi, and T.~T. Yanagida, {\it {Constraint on the
  gravitino mass in hybrid inflation}},  {\em JCAP} {\bf 1012} (2010) 010,
  [\href{http://arxiv.org/abs/1007.5152}{{\tt arXiv:1007.5152}}].

\bibitem{Buchmuller:2014epa}
W.~Buchmuller, V.~Domcke, K.~Kamada, and K.~Schmitz, {\it {Hybrid Inflation in
  the Complex Plane}},  {\em JCAP} {\bf 1407} (2014) 054,
  [\href{http://arxiv.org/abs/1404.1832}{{\tt arXiv:1404.1832}}].

\bibitem{Schmitz:2018nhb}
K.~Schmitz and T.~T. Yanagida, {\it {Axion Isocurvature Perturbations in
  Low-Scale Models of Hybrid Inflation}},  {\em Phys. Rev.} {\bf D98} (2018),
  no.~7 075003, [\href{http://arxiv.org/abs/1806.06056}{{\tt
  arXiv:1806.06056}}].

\bibitem{Buchmuller:2000zm}
W.~Buchmuller, L.~Covi, and D.~Delepine, {\it {Inflation and supersymmetry
  breaking}},  {\em Phys. Lett.} {\bf B491} (2000) 183--189,
  [\href{http://arxiv.org/abs/hep-ph/0006168}{{\tt hep-ph/0006168}}].

\bibitem{Buchmuller:2013lra}
W.~Buchmuller, V.~Domcke, K.~Kamada, and K.~Schmitz, {\it {The Gravitational
  Wave Spectrum from Cosmological $B-L$ Breaking}},  {\em JCAP} {\bf 1310}
  (2013) 003, [\href{http://arxiv.org/abs/1305.3392}{{\tt arXiv:1305.3392}}].

\bibitem{Lazarides:1991wu}
G.~Lazarides and Q.~Shafi, {\it {Origin of matter in the inflationary
  cosmology}},  {\em Phys. Lett.} {\bf B258} (1991) 305--309.

\bibitem{Asaka:1999yd}
T.~Asaka, K.~Hamaguchi, M.~Kawasaki, and T.~Yanagida, {\it {Leptogenesis in
  inflaton decay}},  {\em Phys. Lett.} {\bf B464} (1999) 12--18,
  [\href{http://arxiv.org/abs/hep-ph/9906366}{{\tt hep-ph/9906366}}].

\bibitem{Hamaguchi:2001gw}
K.~Hamaguchi, H.~Murayama, and T.~Yanagida, {\it {Leptogenesis from N dominated
  early universe}},  {\em Phys. Rev.} {\bf D65} (2002) 043512,
  [\href{http://arxiv.org/abs/hep-ph/0109030}{{\tt hep-ph/0109030}}].

\bibitem{Davidson:2002qv}
S.~Davidson and A.~Ibarra, {\it {A Lower bound on the right-handed neutrino
  mass from leptogenesis}},  {\em Phys. Lett.} {\bf B535} (2002) 25--32,
  [\href{http://arxiv.org/abs/hep-ph/0202239}{{\tt hep-ph/0202239}}].

\bibitem{Buchmuller:2005eh}
W.~Buchmuller, R.~D. Peccei, and T.~Yanagida, {\it {Leptogenesis as the origin
  of matter}},  {\em Ann. Rev. Nucl. Part. Sci.} {\bf 55} (2005) 311--355,
  [\href{http://arxiv.org/abs/hep-ph/0502169}{{\tt hep-ph/0502169}}].

\bibitem{Buchmuller:2012bt}
W.~Buchmuller, V.~Domcke, and K.~Schmitz, {\it {WIMP Dark Matter from Gravitino
  Decays and Leptogenesis}},  {\em Phys. Lett.} {\bf B713} (2012) 63--67,
  [\href{http://arxiv.org/abs/1203.0285}{{\tt arXiv:1203.0285}}].

\bibitem{Jeong:2012en}
K.~S. Jeong and F.~Takahashi, {\it {A Gravitino-rich Universe}},  {\em JHEP}
  {\bf 01} (2013) 173, [\href{http://arxiv.org/abs/1210.4077}{{\tt
  arXiv:1210.4077}}].

\bibitem{Kawasaki:2017bqm}
M.~Kawasaki, K.~Kohri, T.~Moroi, and Y.~Takaesu, {\it {Revisiting Big-Bang
  Nucleosynthesis Constraints on Long-Lived Decaying Particles}},  {\em Phys.
  Rev.} {\bf D97} (2018), no.~2 023502,
  [\href{http://arxiv.org/abs/1709.01211}{{\tt arXiv:1709.01211}}].

\bibitem{PhysRevD.98.030001}
{\bf Particle Data Group} Collaboration, M.~Tanabashi, K.~Hagiwara, K.~Hikasa,
  K.~Nakamura, Y.~Sumino, and e.~a. Takahashi, {\it Review of particle
  physics},  {\em Phys. Rev. D} {\bf 98} (Aug, 2018) 030001.

\bibitem{Auclair:2019wcv}
P.~Auclair et~al., {\it {Probing the gravitational wave background from cosmic
  strings with LISA}},  {\em JCAP} {\bf 04} (2020) 034,
  [\href{http://arxiv.org/abs/1909.00819}{{\tt arXiv:1909.00819}}].

\bibitem{Blanco-Pillado:2013qja}
J.~J. Blanco-Pillado, K.~D. Olum, and B.~Shlaer, {\it {The number of cosmic
  string loops}},  {\em Phys. Rev.} {\bf D89} (2014), no.~2 023512,
  [\href{http://arxiv.org/abs/1309.6637}{{\tt arXiv:1309.6637}}].

\bibitem{Smits:2008cf}
R.~Smits, M.~Kramer, B.~Stappers, D.~R. Lorimer, J.~Cordes, and A.~Faulkner,
  {\it {Pulsar searches and timing with the square kilometre array}},  {\em
  Astron. Astrophys.} {\bf 493} (2009) 1161--1170,
  [\href{http://arxiv.org/abs/0811.0211}{{\tt arXiv:0811.0211}}].

\bibitem{Audley:2017drz}
{\bf LISA} Collaboration, P.~Amaro-Seoane et~al., {\it {Laser Interferometer
  Space Antenna}},  \href{http://arxiv.org/abs/1702.00786}{{\tt
  arXiv:1702.00786}}.

\bibitem{LIGOScientific:2019vic}
{\bf LIGO Scientific, Virgo} Collaboration, B.~P. Abbott et~al., {\it {Search
  for the isotropic stochastic background using data from Advanced LIGO’s
  second observing run}},  {\em Phys. Rev.} {\bf D100} (2019), no.~6 061101,
  [\href{http://arxiv.org/abs/1903.02886}{{\tt arXiv:1903.02886}}].

\bibitem{ET}
{\bf Einstein Telescope} Collaboration {\em
  http://www.et-gw.eu/index.php/etsensitivities}.

\bibitem{Verbiest:2016vem}
J.~P.~W. Verbiest et~al., {\it {The International Pulsar Timing Array: First
  Data Release}},  {\em Mon. Not. Roy. Astron. Soc.} {\bf 458} (2016), no.~2
  1267--1288, [\href{http://arxiv.org/abs/1602.03640}{{\tt arXiv:1602.03640}}].

\bibitem{Shannon:2015ect}
R.~M. Shannon et~al., {\it {Gravitational waves from binary supermassive black
  holes missing in pulsar observations}},  {\em Science} {\bf 349} (2015),
  no.~6255 1522--1525, [\href{http://arxiv.org/abs/1509.07320}{{\tt
  arXiv:1509.07320}}].

\bibitem{Ade:2015xua}
{\bf Planck} Collaboration, P.~A.~R. Ade et~al., {\it {Planck 2015 results.
  XIII. Cosmological parameters}},  {\em Astron. Astrophys.} {\bf 594} (2016)
  A13, [\href{http://arxiv.org/abs/1502.01589}{{\tt arXiv:1502.01589}}].

\bibitem{Thrane:2013oya}
E.~Thrane and J.~D. Romano, {\it {Sensitivity curves for searches for
  gravitational-wave backgrounds}},  {\em Phys. Rev.} {\bf D88} (2013), no.~12
  124032, [\href{http://arxiv.org/abs/1310.5300}{{\tt arXiv:1310.5300}}].

\end{thebibliography}\endgroup

%%%%%%%%%%%%%%%%%%%%%%%%%%%%%%%%%%%%%%%%%%%%%%%%%%%%%%%%%%%%%%%%%%%%%%%%%%%%%%%%%%%%%%%%%%%%%%%%%%%%

\end{document}